\documentclass[10pt,letterpaper]{article}
\usepackage[top=0.85in,left=2.75in,footskip=0.75in]{geometry}

\usepackage{changepage}

\usepackage[utf8]{inputenc}

\usepackage{textcomp,marvosym}

\usepackage{fixltx2e}

\usepackage{amsmath,amssymb,amsfonts}

\usepackage{cite}

\usepackage{nameref,hyperref}

\usepackage{microtype}
\DisableLigatures[f]{encoding = *, family = * }

\usepackage{rotating}

\usepackage{setspace} 
\raggedright
\usepackage[aboveskip=1pt,labelfont=bf,labelsep=period,justification=raggedright,singlelinecheck=off]{caption}

\makeatletter
\renewcommand{\@biblabel}[1]{\quad#1.}
\makeatother

\date{}

\usepackage{lastpage,fancyhdr,graphicx}
\usepackage{epstopdf}
\fancyheadoffset[L]{2.25in}
\fancyfootoffset[L]{2.25in}

\usepackage{graphicx}

\newcommand{\tl}{\tau_{\rm lag}}
\newcommand{\gm}{g_{\rm max}}

\begin{document}
\vspace*{0.35in}

\begin{flushleft}
{\Large
\textbf\newline{\bf Growth of bacteria in 3-d colonies}
}
\newline
\\
Xinxian Shao\textsuperscript{1}, Andrew Mugler\textsuperscript{1,2},
Justin Kim\textsuperscript{3}, Ha Jun Jeong\textsuperscript{3}, Bruce
Levin\textsuperscript{3}, Ilya Nemenman\textsuperscript{1,3},
\\
\bigskip {\bf 1} Department of Physics, Emory University, Atlanta, GA
30322, USA
\\
{\bf 2} Department of Physics and Astronomy, Purdue University, West Lafayette, IN
47907, USA
\\
{\bf 3} Department of Biology, Emory University, Atlanta, GA 30322,
USA
\\
\bigskip * ilya.nemenman@emory.edu
\end{flushleft}

\section*{Abstract}

The dynamics of growth of bacterial populations has been extensively
studied for planktonic cells in well-agitated liquid culture, in
which all cells have equal access to nutrients.  In the real world,
bacteria are more likely to live in physically structured habitats as
colonies, within which individual cells vary in their access to
nutrients.  The dynamics of bacterial growth in such conditions is
poorly understood, and, unlike that for liquid culture, there is not a
standard broadly used mathematical model for bacterial populations
growing in colonies in three dimensions (3-d). By extending the
classic Monod model of resource-limited population growth to allow for
spatial heterogeneity in the bacterial access to nutrients, we develop
a 3-d model of colonies, in which bacteria consume diffusing nutrients
in their vicinity. By following the changes in density of {\em
  E.~coli} in liquid and embedded in glucose-limited soft agar, we
evaluate the fit of this model to experimental data.  The model
accounts for the experimentally observed presence of a
sub-exponential, diffusion-limited growth regime in colonies, which is
absent in liquid cultures. The model predicts and our experiments
confirm that, as a consequence of inter-colony competition for the
diffusing nutrients and of cell death, there is a non-monotonic
relationship between total number of colonies within the habitat and
the total number of individual cells in all of these colonies. This
combined theoretical-experimental study reveals that, within 3-d
colonies, {\em E.~coli} cells are loosely packed, and colonies produce
about 2.5 times as many cells as the liquid culture from the same
amount of nutrients. Our model provides a baseline description of
bacterial growth in 3-d, deviations from which can be used to identify
phenotypic heterogeneities and inter-cellular interactions that
further contribute to the structure of bacterial communities.

\section*{Author Summary}
It is convenient for theoreticians as well as experimentalists to
maintain the fiction that bacteria exist as planktonic cells in
well-mixed liquid cultures, all with equal access to nutrients,
wastes, toxins, antibiotics, bacterial viruses, and each
other. However, in the real world, bacteria are more often found in
physically structured, spatially heterogeneous habitats as colonies
and micro-colonies. While one can experimentally explore the
population and evolutionary dynamics of bacteria in such physically
structured habitats, there is dearth of mathematical models to
generate hypotheses for and to interpret results of these
experiments. As a step towards the construction of a theory of the
population dynamics of bacteria in physical structured habitats, we
develop and experientially explore the simplest such model of the
dynamics of bacterial growth in 3-d structured colonies. 

\section*{Introduction}
In 1942, Jacques Monod developed a mathematical model of bacterial
growth in a liquid culture, where the bacterial cells and nutrient
molecules were homogeneously distributed
\cite{monod1942recherches,monod1949}.  A simple ordinary differential
equation was accurate enough to account for the exponential growth of
bacteria and their ascent into stationary phase following the
exhaustion of the limiting resource.  The model has proven to be
long-lived since most experimental studies of the population dynamics
of bacteria are in liquid
culture\cite{kubitschek1970introduction,StewartLevin73}.  In contrast,
outside the tubes, flasks, and chemostats of laboratory culture,
bacteria most commonly live as physically structured habitats as
colonies or microcolonies. Such colonies are heterogeneous; at a
minimum, their cells vary in their access to nutrients depending on
their position within the colony and thereby divide at different
rates.

The majority of research directed at understanding structured bacterial
population growth has been confined to two dimensional (2-d) surfaces
\cite{BenJacob:1994db,BenJacob:1995vu,softagar,fractal,Liu:2011ei,wrinkle},
including studying the interplay of evolution and the physical
structure \cite{Wei:2011ca,Kim:2014ie}, or analyzing effects of
mechanical interactions in an expanding colony
\cite{Mertz2012,Farrell:2013dg,Ghosh2015}. However, diffusion in two
dimensions is different from three, making it easier to form
diffusion-limited instabilities
\cite{Witten:1981ub,BenJacob:1994db,Family:1995vt}. In 3-d, work has
been done to understand nutrient shielding of the interior of a colony
by the microbes on the surface, treating them as individual agents
\cite{yeastcluster}.  However, we are not aware of 3-d models of
colony growth that account for the spatially varying density of
nutrients and bacteria, explain the observed experimental
phenomenology of bacterial growth in such colonies, and do so in a
relatively simple coarse-grained (PDE) Monod-style manner, rather than
relying on complex agent simulations of individual bacteria

Here we develop such a model that treats the growth rate heterogeneity
due to the non-uniform nutrient distribution produced
self-consistently by consumption of a nutrient by the bacteria.  We
explore its fit experimentally with the growth of {\em E.~coli}
maintained and growing as colonies embedded in 3-d matrix of soft agar
with an initially uniform spatial distribution of a limiting carbon
source, glucose (Fig.~\ref{image}). We compare dynamics of growth of
bacteria in colonies with that of planktonic cells in liquid culture
with the same concentration of limiting glucose. In our model, we
assume that the colony is essentially unconstrained by the soft agar
and is free to expand, and the bacteria within them are
non-motile. This combined theoretical-experimental study reveals two
surprising features of bacterial populations growing as colonies: (i)
the bacteria within these structures exist as loosely packed viable
cells, and (ii) the viable cell densities of bacteria produced in
colonies is more than two-fold greater than that in liquid cultures
with the same concentration of the limiting glucose.

\section*{Results}

\subsection*{Experimental characterization of bacterial growth}

\begin{figure}[t]
\includegraphics[width=1\linewidth]{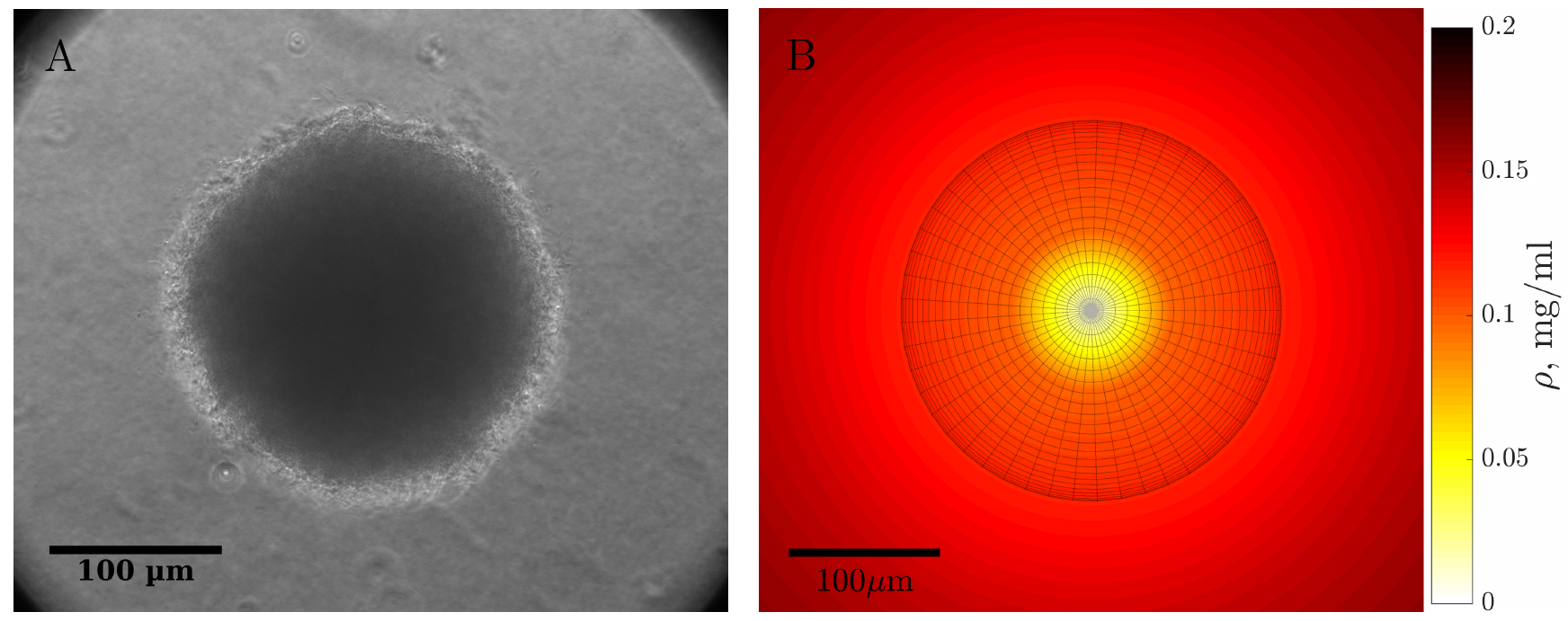}
  \caption{{\bf 3-d colony growth.} (A) Photograph of a representative {\em
      E.~coli} colony inside 3-d agar at 22 hrs post inoculation. (B)
    A growing colony at 22 hrs as simulated using our mathematical
    model. Heatmap shows the spherically symmetric nutrient
    concentration, and the meshgrid sphere represents the colony. At
    this time, the nutrient at the center of the colony is fully
    consumed.  Since the growth rate depends on local nutrient
    concentration, the cells at the center of the colony are not growing
    anymore. \label{image}}
\end{figure}

We use population growth of {\em E.~coli} in minimal medium as the
basis for developing the model.  To control the amount of nutrients
available to the bacteria, we use glucose at the initial concentration
$0.2$ mg/ml, at which it limits the stationary phase density of {\em
  E.~coli} produced as planktonic cells and as colonies in soft
agar. We grow bacteria either in liquid cultures or as
three-dimensional colonies embedded in soft agar
(Fig.~\ref{image}). Unless otherwise noted, 3-d colonies are grown in
$3$ ml of soft agar, inoculated with approximately $50$
bacteria/ml. Under these conditions, each colony has an access to a
nutrient subvolume of $v\sim1/50$ ml, or, on average, a nutrient
sphere of radius $R=(3v/4\pi)^{1/3}\approx 1.7$ mm. For the liquid and
the 3-d growth, we estimate the density of viable cells, $N(t)$, at
different times diluting and plating and then counting the number of
resulting colonies (colony-forming units, or CFU), see {\em Methods}
for details. For each time point, we obtain 6 independent replicates
of CFU density estimates, and each experiment was replicated at least
3 times.

The results of these population growth experiments are shown in
Fig.~\ref{datacompiled} (data points).  In liquid, the density of the
population increases exponentially, and then abruptly stops and begins
to decline at a low rate, presumably because the bacteria consume the
available glucose and enter the stationary phase, at which time the
rate of cell mortality exceeds that of division. In contrast, in 3-d
colonies, the exponential growth and the stationary phase are
separated by a gradual decline in the net rate of growth.  We expect
that this is because the growth of the population here is limited by
the speed with which diffusion brings glucose from the periphery of
the available nutrient volume to the colony, where it is consumed by
the bacteria.  Surprisingly, the maximum density of bacteria growing
as colonies is substantially greater than that in liquid, despite the
concentration of the limiting glucose being equal for liquid and the
soft agar. To understand these findings quantitatively, we now develop
a simple (minimalist) mathematical model of resource-limited bacterial growth
in liquid and as spatially structured colonies.

\subsection*{Minimal model of resource-limited bacterial growth}

\begin{figure}[t]
\includegraphics[width=.7\linewidth]{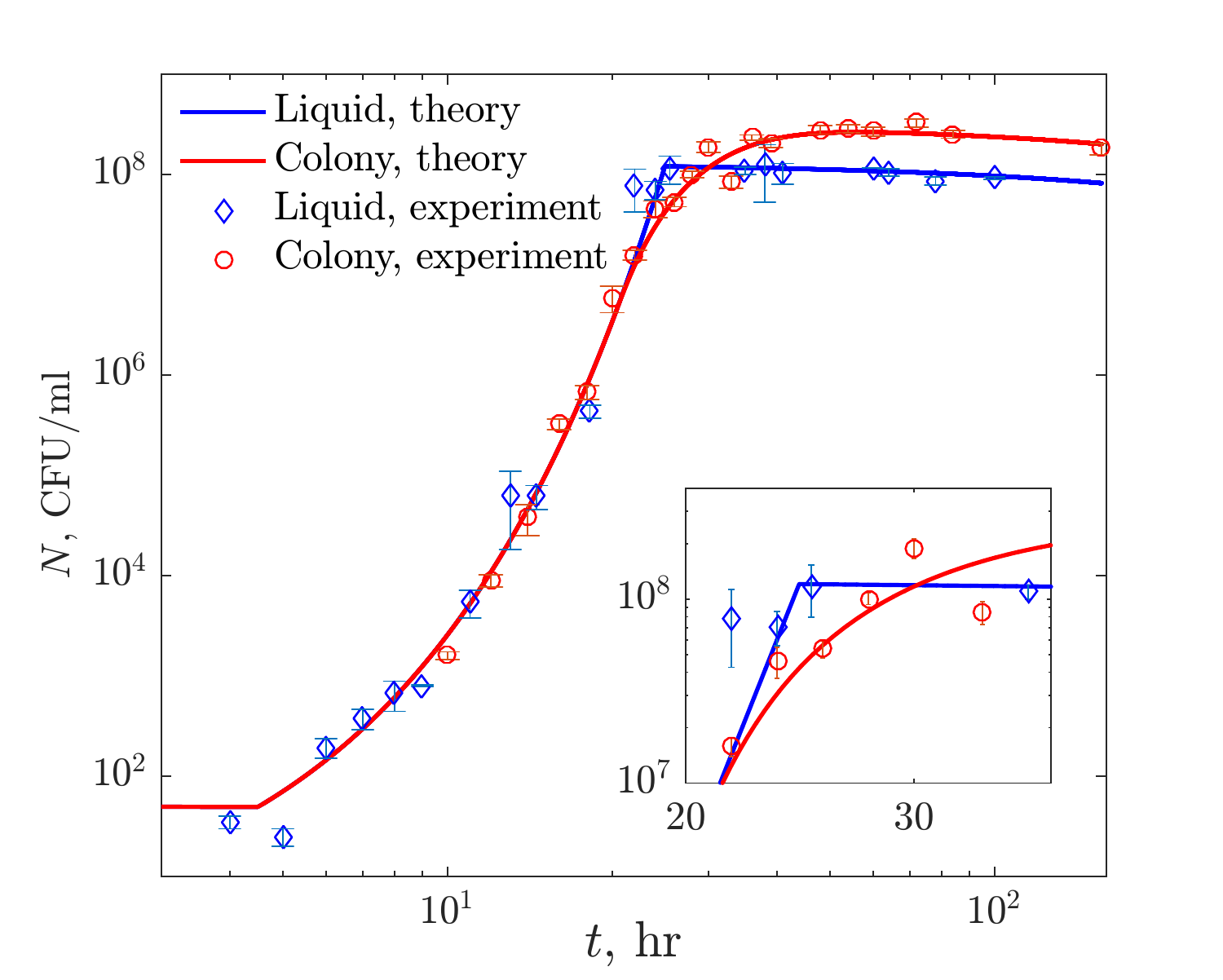}
\caption{{\bf {\em E.~coli} population dynamics.} Experimental data,
  averaged over all experiments (symbols, error bars are s.~e.~m.),
  are compared with the fits of the mathematical model we developed
  (solid lines). For clarity, uncertainty of the numerical predictions
  is omitted and is shown instead in Fig.~\ref{80cband}. Liquid
  cultures switch abruptly from the exponential growth to the
  saturation, and then decay slowly. In contrast, 3-d colonies
  gradually slow down before saturating (see Inset) at a population
  size larger than that in the liquid, and then
  decay. \label{datacompiled}}
\end{figure}

Our liquid culture model of bacterial growth is a variant of that of
Monod \cite{monod1949}.  In this model, all bacteria have the same
resource (glucose) concentration dependent growth rate
$g(\rho)=\gm\rho/(\rho+K)$, where $\rho$ is the concentration of
glucose, $\gm$ is the maximum growth rate, and $K$, the Monod
constant, is the concentration of the resource when the growth rate is
half its maximum value $\gm/2$.  With these parameters, the rate of
change of the density of the bacterial population $n = N/v$ is given
by
\begin{align}
  \frac{dn}{dt}&=n\,\Theta(t-\tl)\frac{\gm\,\rho}{\rho+K}-nm\label{qg2},\\
  \frac{d\rho}{dt}&=-\frac{1}{a_{\rm l}}n\,\Theta(t-\tl)\frac{\gm\,\rho}{\rho+K}\label{qg3}.
\end{align} 
Here $v$ is the volume of the liquid where the culture grows, and
$a_{\rm l}$ is the liquid {\em yield}, which measures the number of
bacteria produced by a microgram of the nutrient. Further,
$\Theta(t-\tl)$ is the Heaviside $\Theta$-function, which is equal to zero for
$t<\tl$, and to unity otherwise. It represents the lag phase before
the growth starts after a transfer to a new environment. Note that
Eqs.\ \eqref{qg2} and \eqref{qg3} differ slightly from the standard Monod
model. Specifically, we added a small constant death rate $m$ to
account for the decrease of the population in the liquid culture after
the saturation (Fig.~\ref{datacompiled}). Thus the population has a zero
net growth at a critical nutrient density of
$\rho_{\rm m}= mK/(\gm-m)$, which represents the minimum nutrient
concentration needed to sustain life without growth \cite{minsu}.

We fit the five growth parameters ($g_{\rm max}$, $K$, $a_{\rm l}$,
$\tl$, and $m$) to the experimental data using nonlinear least squares
fitting, and estimate uncertainties of the fit using bootstrapping
(see {\em Materials and Methods}, and also Table~\ref{fittedpars}). As
seen in Fig.~\ref{datacompiled} (blue curve), after the lag phase, the
population increases exponentially before it saturates abruptly when
all the cells in the colony run out of food at the same time. The
excellent agreement between the experiments and the model is
encouraging. It allows us to use the Monod model with death as the
basis for 3-d studies.

To develop the minimal model of 3-d growth, we assume that the
bacteria within the colony are physiologically identical, but
depending on their position, vary in their access to the diffusing
carbon source. Thus all cells grow according to the Monod model,
differing only by the local availability of the limiting nutrient, glucose,
$\rho(x,y,z)$.  We assume that soft agar is too soft to provide
mechanical resistance to the colony,  but 
sufficiently dense to keep cells from moving. As the cells
divide, the colony expands symmetrically as a sphere, keeping a
constant cell density per volume of the space occupied by the colony.
This leads to the following equations for the spherically-symmetric
local number density of cells $n(r,t)$ and nutrient concentration
$\rho(r,t)$:
\begin{align}
  \frac{\partial n(r,t)}{\partial t}&=n(r,t)\,\left[\Theta(t-\tl)\frac{\gm\,\rho(r,t)}{\rho(r,t)+K}-m\right]\label{qg4},\\
  \frac{\partial \rho(r,t)}{\partial t}&=D\nabla^2\rho
  -\frac{1}{a_{\rm c}}n(r,t)\,\Theta(t-\tl)\frac{\gm\,\rho(r,t)}{\rho(r,t)+K}\label{qg5},
\end{align}
with the initial uniform spatial concentration of the nutrient
$\rho(r,0) = \rho_{0}$ at time $t=0$, and a single bacterium starting
the colony at $r=0$.  In these equations, $D$ is the nutrient
(glucose) diffusion coefficient.  Further, we allow for the yield in
the colony $a_{\rm c}$ to be different from the liquid yield
$a_{\rm l}$ to account for the different saturation values in
Fig.~\ref{datacompiled}, as further discussed below. Importantly,
since the agar is more than 99\% liquid, the four other growth
parameters $\gm$, $K$, $\tl$, and $m$ are taken to be the same in both
media.

To keep the colony at the same fixed cellular packing density $\mu$,
we require that the overall increase in cell number leads to the
proportionate growth of the colony radius $r_{\rm c}$, so that
$N\equiv 4\pi \int dr\, r^2 n(r,t)=(4/3) \pi r_{\rm c}^3 \mu$. Thus at
each point in time, we impose the condition that
\begin{equation}
n(r,t)=\left\{
\begin{array}{ll}
  \mu,&  0<r\leq r_{\rm c}=(3N/4\pi \mu)^{1/3},\\
  0,&r_{\rm c}<r\le R,\\
\end{array}
\right.
\label{nrt}
\end{equation}
where $R=(3v/4\pi)^{1/3}$ is the radius of the nutrient subvolume
accessible to the colony. To reconcile Eqs.~\eqref{qg4} and
\eqref{nrt}, we say that all new growth is immediately transferred to
the colony edge, $r_{\rm c}(t)$, while the death results in a decrease
in the cell density locally (see {\em Materials and Methods} for
description of the algorithm for simulating this growth model).

\begin{figure}[t]
\includegraphics[width=0.5\linewidth]{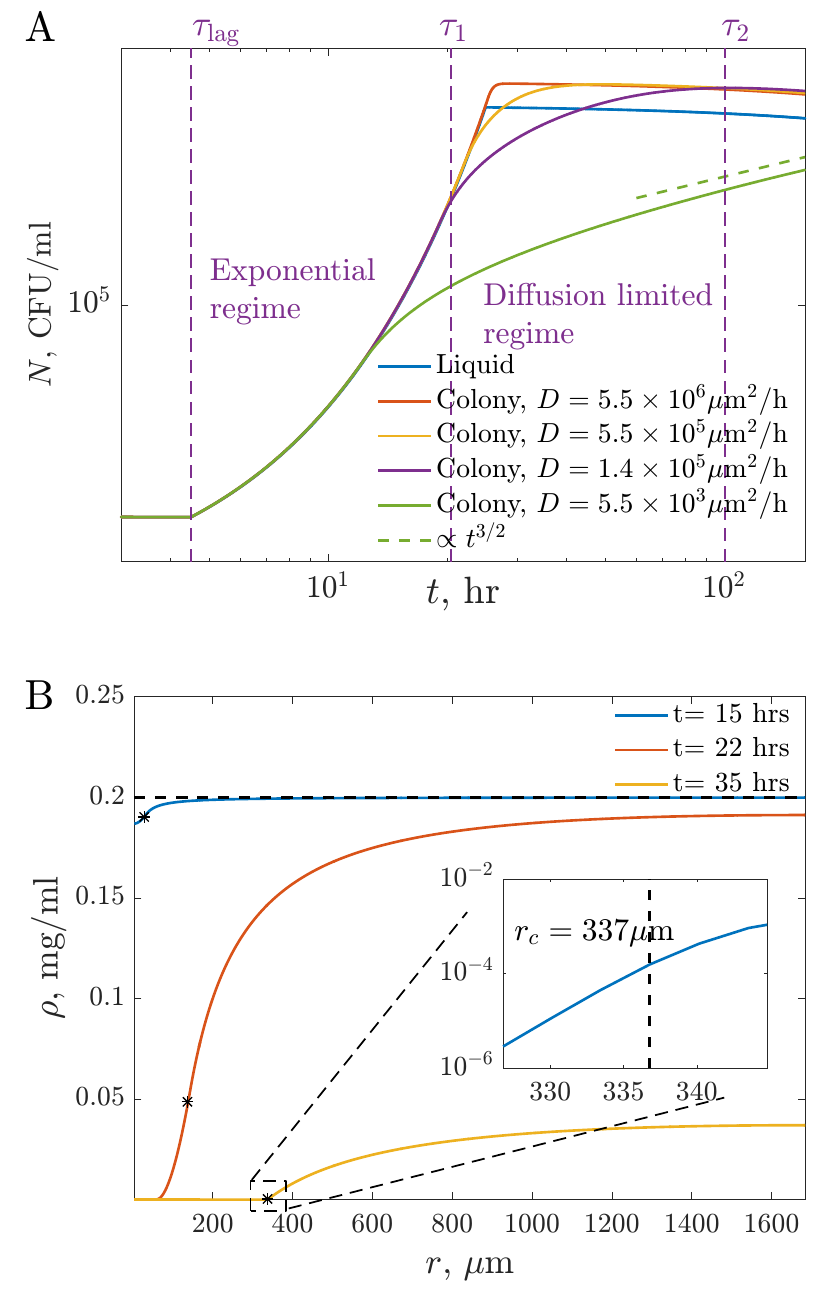}
\caption{{\bf Mathematical model predictions.} (A) Population growth
  in liquid culture and in 3-d colonies. The growth parameters are
  chosen as best fit values for our experimental data (see
  Tbl.~\ref{fittedpars}), except for $D$, which we vary to illustrate
  different growth regimes. The diffusion-limited regime in the limit
  of small $D$ is consistent with the prediction $N\propto
  t^{3/2}$.
  The time scales $\tau_i$ are illustrated for $D=1.4\times 10^5$
  $\mu$m$^2$/hr.  (B) Profile of the nutrient concentration in space
  at different times using the same parameters as above and 
  $D=5.5\times 10^5$ $\mu$m$^2$/h, as in Tbl.~\ref{fittedpars}. The
  edge of the colony is illustrated by stars on each curve. The inset
  shows that the concentration decreases exponentially at the colony
  edge in the diffusion-limited growth regime. The penetration depth
  is about $3$ $\mu$m.  
  \label{modelsolutions}}
\end{figure}

To illustrate the behavior of this 3-d model of bacterial growth as
colonies, we plot numerical solutions of Eqs.~\eqref{qg4}-\eqref{nrt}
for different values of the nutrient diffusion coefficient in
Fig.~\ref{modelsolutions} (A).  Especially at small $D$, two different
growth regimes are clearly visible after the lag but before the
ultimate saturation and the slow cell death. The first is the fast
{\em exponential growth} based on local, immediately accessible
resources. This regime is indistinguishable from the growth in liquid.
When the local nutrients are depleted at a certain time $\tau_1$
following the start of the growth at $\tl$, new nutrients must be
brought from afar by diffusion.  This is slow, resulting in a slower
{\em diffusion-limited growth} regime. Here the overall colony growth
rate is an average over cells growing at different rates due to
different concentrations of the locally accessible nutrient. Our
simulations suggest that, in this regime, the nutrient concentration
at the colony edge decays exponentially fast, in agreement with
Ref.~\cite{yeastcluster}, cf.~Fig.~\ref{modelsolutions} (B). The
nutrient penetration depth is only a few $\mu$m, or a few cell
layers. Therefore, in the diffusion-limited regime, there are,
essentially, no nutrients deep inside a colony, and only cells at the
periphery can grow. In the absence of resource storage
~\cite{saint2014massive}, nutrient sharing from the outer cells, or
cannibalism (we model none of these), interior cells would not grow at
all and will eventually die. The diffusion-limited growth regime
finally ends with {\em saturation} and {\em slow death} when most of
the nutrients in the accessible subvolume are depleted at time
$\tau_2$ after $\tl$. The onset of the saturation takes longer than in
liquid since small (but larger than $\rho_{\rm m}$) amounts of the
nutrient linger at the far edges of the nutrient subvolume for a long
time.

Analytical expressions for $\tau_1$, $\tau_2$, and the growth dynamics
can be obtained from the following arguments.  First, in the
exponential growth regime, the population grows as $N\sim e^{\gm t}$.
This requires $e^{\gm t}/a_{\rm c}$ of the nutrient mass, which must
come from the volume immediately accessible by diffusion, equal to
$\sim \rho_0 (\sqrt{Dt})^{3}$. Equating the two expressions gives, to
the leading order,
$\tau_1\sim \gm^{-1}\log[\rho_0 a_{\rm c} (D/\gm)^{3/2}]$.  When
local resources are exhausted, growth is limited by nutrients
diffusing in from the volume $\sim (\sqrt{Dt})^{3}$. However, because
the encounter probability for a 3-d random walk is less than one
\cite{redner}, most of the nutrient molecules coming from afar will
not be immediately absorbed. In fact, since the box-counting dimension
of a diffusive process is two, only
$\sim \rho_0 (\sqrt{Dt})^{2}r_{\rm c}$ nutrient molecules will be
captured in time $t$, resulting in
$N\sim \rho_0 Dtr_{\rm c}a_{\rm c}$. On the other hand, the radius of
the colony grows as $r_{\rm c}= (3/4\pi)^{1/3}
(N/\mu)^{1/3}$.
Combining these expressions gives
$N \sim [(a_{\rm c}\rho_0D)^3/\mu]^{1/2} t^{3/2}$ in the
diffusion-limited regime. Finally, the total amount of nutrients
available to the colony is $\sim \rho_0R^3$, and so the
diffusion-limited growth will saturate, and the cells will start dying
with the rate of $m$ when the colony grows to
$N \sim a_{\rm c} \rho_0R^3$, which occurs at
$\tau_2\sim (\mu/a_{\rm c}\rho_0)^{1/3}R^2/D$.  Altogether, we find
\begin{equation}
N\sim\left\{
\begin{array}{lll}
  {\rm const}, & t<\tl,\\
  e^{\gm t},&  t-\tl\ll\tau_1\sim \frac{\log\left[\rho_0a_{\rm c} \left(\frac{D}{\gm}\right)^{3/2}\right]}{\gm},\\
  \left[\frac{(a_{\rm c}\rho_0D)^3}{\mu}\right]^{\frac{1}{2}}
  t^{3/2},&\tau_1\ll t-\tl\ll\tau_2\sim \left(\frac{\mu}{a_{\rm c}\rho_0}\right)^{\frac{1}{3}}\frac{R^2}{D},\\
   a_{\rm c} \rho_0R^3 e^{-mt},& \tau_2\ll t-\tl,
\end{array}
\right.
\label{qg6}
\end{equation}
These analytical estimates are supported by the numerical solutions in
Fig.~\ref{modelsolutions}(A).

We note that in one or two dimensions, the diffusion limited growth
would scale as $N\propto t^{d/2}$ for dimension $d$, independently of
the (small) colony radius, or even for a point colony, since the
random walk encounter probability there is one \cite{redner}. In
contrast, our three-dimensional result depends critically on knowing
how the radius of the colony scales with the number of growing
bacteria. In particular, here we cannot model the colony as a
point-like object. Thus the exponent of the power law scaling is not
universal in 3-d, and it may change for heterogeneous colonies with
varying cell size and cell density.

\subsection*{Experimental tests of the minimal model of bacterial growth}
\begin{figure}[t]
\includegraphics[width=0.5\linewidth]{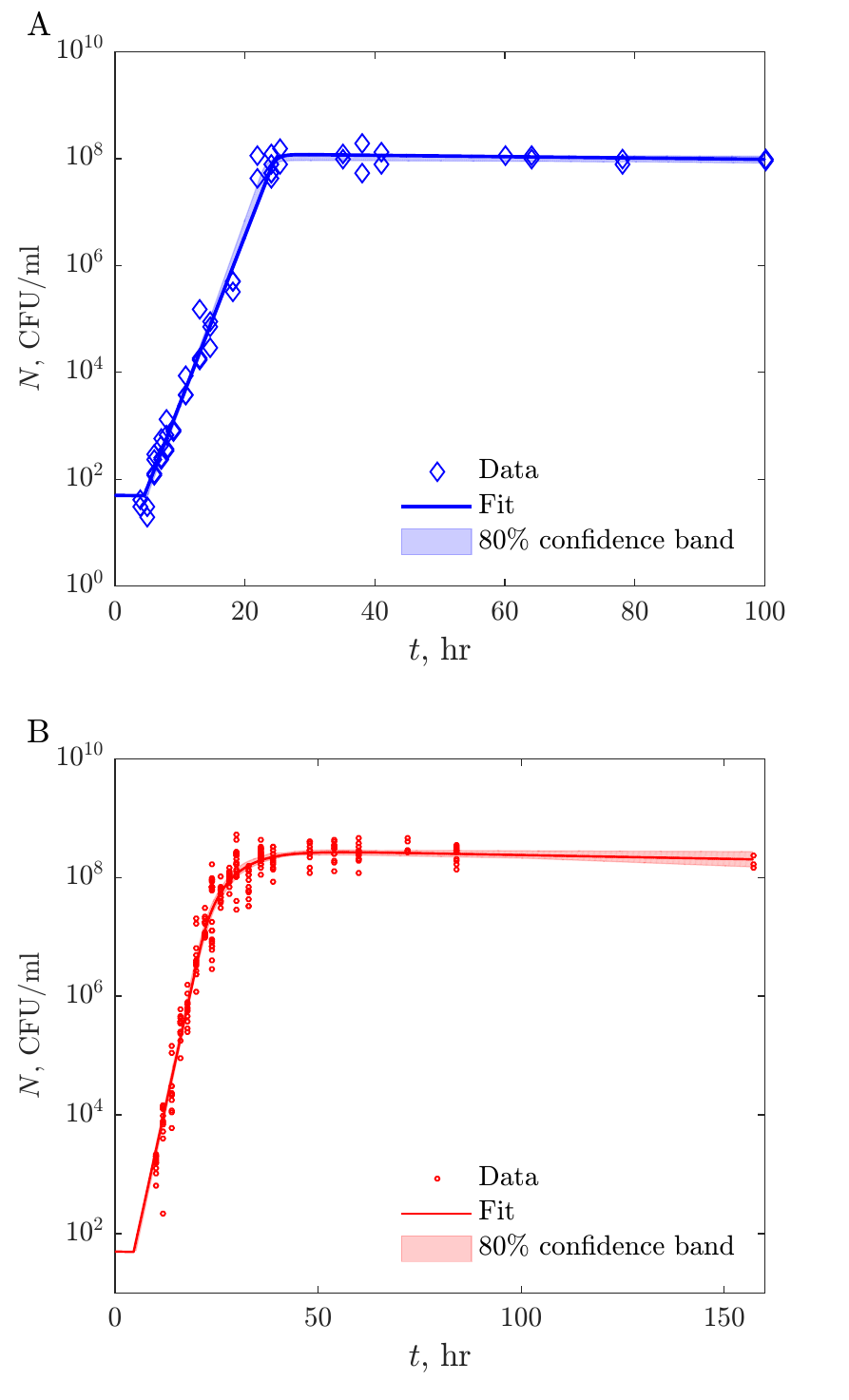}
\caption{{\bf Fitting models to data}. {\em A}: Liquid growth
  model (solid blue line) fitted to all of the experimental data we
  have collected (blue diamonds). 80\% confidence intervals around the
  best-fit predictions are shown by light blue shaded bands
  (established by bootstrapping, with 1000 resamplings). {\em B}:
  same, but for 3-d colony growth. Red circles, solid red line, and
  light red band correspond to the data, the best fit, and the 80\%
  confidence intervals (from 30 resamplings). \label{80cband}}
\end{figure}

To determine the extent to which our minimal model accounts for the
dynamics of growth of bacteria in colonies, we fit the model to data
using nonlinear least squares fitting, similar to the liquid case.  We
keep the parameters $a_{\rm l}$, $K$, $\gm$, $m$, and $\tl$ equal to
the values inferred for liquid, and only optimize $D$, $\mu$, and
$a_{\rm c}$ for the 3-d culture data.  See {\em Materials and Methods}
for the details of the fits, including estimation of the prediction
uncertainty using bootstrapping. Table~\ref{fittedpars} shows fitted
parameter values with the corresponding nominal values from the
literature. The fitted parameters are consistent with the nominal
values where the latter are known. Further, the best fit curve shows
an excellent agreement with data (cf.~Fig.\ \ref{datacompiled}, red),
and the prediction confidence bands are very narrow
(cf.~Fig.~\ref{80cband}). This suggests that nutrient diffusion and
the ensuing geometric heterogeneity of growth are sufficient to
explain the population dynamics of these {\em E.~coli} colonies in 3-d
at our experimental precision, and consideration of additional
phenotypic inhomogeneities is not needed.

Our analysis also provides estimates of two previously unknown
parameters, $\mu$ (packing density) and $a_{\rm c}$ (yield in 3-d
colonies). The inferred packing density is $\mu = 3.0 \cdot 10^{-2}$
CFU/$\mu$m$^3$, with the 80\% confidence interval of
$[1.7, 4.2]\cdot 10^{-2}$ CFU/$\mu$m$^3$ . Since an {\em E.~coli} cell
has a volume of between $0.5$ and $2$ $\mu$m$^3$
\cite{Pierucci1978,Taheri2014}, this suggests that only about
$\sim 3$\% of all space in a colony is occupied by viable cells. This
is a surprising finding, and it requires an independent
corroboration. Towards this end, we measure radii of large colonies
and calculate their packing densities by diving colony volumes by the
average CFUs per colony. This gives $\mu=1.5\pm 0.08 \times10^{-2}$
CFU/$\mu{\rm m}^3$, consistent with our estimation of $\mu$ from the
fitted growth model. In other words, in our experiments, viable {\em
  E.~coli} cells like to keep their distance from each other.

The second inferred parameter is $a_{\rm c}$. We find that the yield
as measured by the ratio of the CFU estimated stationary phase density
and the quantity of glucose in 3-d is 2 to 3 times higher than that in
liquid culture, $a_{\rm c}>a_{\rm l}$ (cf.~Table
\ref{fittedpars}). This implies that, at saturation, colonies produce
more CFUs than liquid cultures, which is directly apparent from Fig.\
\ref{datacompiled}. This is a surprising result, since in the colony
the bacteria grow more slowly and there is more time for cell
death. Nonetheless, similar results have been reported for colonies
growing on surfaces \cite{SSS}. Here this effect is likely a direct
consequence of the growth dynamics during the diffusion-limited
regime. Indeed, {\em E.~coli} cells growing at a rate of $>1$
hr$^{-1}$ grow to be $2$ to $3$ times larger than cells growing at a
rate of $<0.1$ hr$^{-1}$ \cite{cellsize}. While the diffusion limited
regime lasts only for a few hours (cf.~Fig.~\ref{datacompiled}), more
than 90\% of all cells emerge at that time, so that the majority of
cells in the colony are smaller than in liquid, yielding more cells
from the same nutrient amount.

\begin{figure}[t]
\includegraphics[width=.7\linewidth]{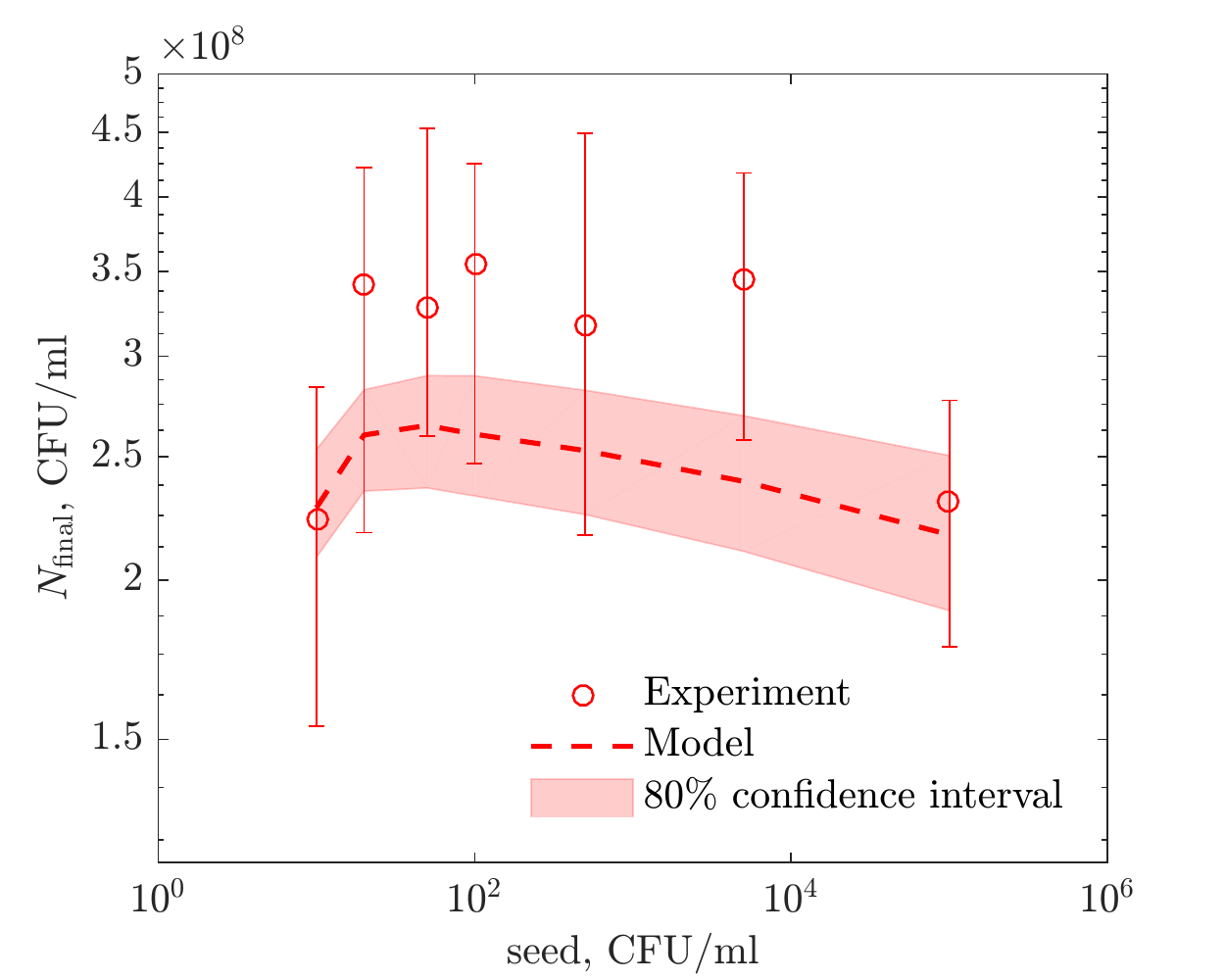}
\caption{{\bf Dependence of the population size on the inoculation
    density.} Colony cultures inoculated with different cell densities
  grow to different population sizes. Circles are experiment data
  measured at 74 hr post inoculation, and error bars are s.~e.~m. The
  best-fit 3-d bacterial growth model reproduces these data within
  experimental error bars and computational confidence interval,
  without additional fitting. \label{multiseed}}
\end{figure}

As an independent test of the developed 3-d growth model, we use it to
predict results of experiments distinct from those used for fitting
the model.  Specifically, we investigate how the population size
depends on the density of bacteria used to inoculate the soft agar. At
a long measurement time (72 hrs), our model predicts a non-monotonic
dependence of the population size on the inoculation density
(cf.~Fig.~\ref{multiseed}, dashed line). This is because, at very low
densities, each colony has access to a large nutrient subvolume, and
the colony cannot clear this subvolume by diffusion in just 72 hrs. As
a result, at the end of the experiment, there are still nutrients in
the media, and the colony does not reach its maximum size. In
contrast, at very high inoculating densities, colonies rapidly exhaust
their small available nutrient subvolumes, the cell death becomes
important throughout much of the experiment duration, and the
population is smaller again. Thus the population reaches its maximum
at intermediate densities, where these two effects balance. We test
this prediction by experimentally measuring population sizes at 72 hrs
for {\em E.~coli} growing in soft agar at inoculums varying from
$10^1$ to $10^5$ cells/ml As seen in Fig.~\ref{multiseed}, the
experimental data agree with the prediction within errors and, in
particular, exhibit the expected non-monotonicity.  We emphasize that
no additional fitting was done for this figure, and yet the agreement
between the experiment and the theory is very good.

\begin{table}[!ht]
\begin{adjustwidth}{-2.25in}{0in} 
\caption{
{\bf Fitted parameters }}
\begin{tabular}{|l|l|l|l|l|l|}
  \hline
  \textbf{Name} & \textbf{Description}  & \textbf{Literature}
  &{\bf References}&{\bf Fitted value}& \textbf{80\% confidence}\\
   &   & \textbf{values}
  &&& \textbf{interval}\\
  \hline
  $g_{\rm max}$ 
                & maximum growth rate, ${\rm hr}^{-1}$ 
                                        & $[0.52, 0.83]$&
                                          \cite{Pierucci1978,Pelletier2012,Fuchslin2012}
  & $0.73$ & $[0.56, 0.89]$ \\  \hline
  $K$ 
                & half-saturation constant, ${\rm \mu g}/{\rm L}$ 
                                        & $35$ &\cite{Fuchslin2012}
  & $122$ & $[19.5, 783]$ \\   \hline
  $a_{\rm l}$ 
                & yield in liquid, $10^{6}$ CFU/$\mu$g\ glucose
                                        & $[0.5,1.2]$ &\cite{Levin1972,Chao1981}
  & $0.61$ 
                     & $[0.54, 0.67]$\\  \hline
  $\tl$ 
                & lag phase duration, ${\rm hr}$ 
                                        & $[2,5]$&\cite{buchanan1992effect}
  & $4.5$
                     & $[4.3, 4.7]$ \\  \hline 
  $m$ 
                & death rate, ${\rm hr^{-1}}$ 
                                        & $[0.0049, 0.018]$ & \cite{minsu,Fuchslin2012} 
  & $0.0029$ & $[7.7\times 10^{-4}, 0.011]$ \\  \hline
  $a_{\rm c}$ 
                & yield in 3-d colony, $10^{6}$  CFU/$\mu$g  glucose
                                        & N/A &
  & $1.50$
                     & $[1.36, 1.63]$\\  \hline 
  $D$ 
                & glucose diffusion in $0.35\%$ agar, $\mu$m$^{2}$/hr
                                        & $1.8\times 10^6$ & \cite{dcglu}
  &	$0.55\times 10^6$
                     & $[0.21, 0.89]\times 10^6$ \\  \hline 
  $\mu$ 
                & packing density, ${\rm CFUs}/\mu{\rm m}^3$ 
                                        & N/A &
  & $2.98\times10^{-2}$
                     & $[1.74, 4.22]\times10^{-2}$\\
  \hline
\end{tabular}
\label{fittedpars}
\end{adjustwidth}
\end{table}

\section*{Discussion}

To our knowledge, the model developed here is the first continuous,
rather than agent-based, model to explicitly study bacterial growth as
colonies. We consider this the minimal model because it assumes that
the availability of nutrients (a carbon source) is the sole factor
determining the rate of cell division within colonies. In reality, the
cellular growth, division, and death rates would also depend on
cell-to-cell interactions of various sorts, on the enrichment and
deterioration of the environment due to the buildup of secondary
metabolites and waste, on cell-environment mechanical interactions,
and on diverse cellular phenotypic commitments. The model we developed and
experimentally tested here only accounts for the spatial
heterogeneity in access to the diffusing nutrient and assumes no such
additional effects \cite{Asally:2012gj,Wentland1996,saint2014massive}.

Nevertheless, despite these limitations, with only five parameters
describing the growth in liquid, and only three additional parameters
specific to the 3-d colony growth, this model provides an impressively
accurate description of growth of populations of {\em E.\ coli} as
colonies in soft agar as well as planktonic cells in liquid. Unlike
the anticipated and observed nearly precipitous termination of growth
in liquid culture as nutrients become depleted, our 3-d model accounts
for the experimentally observed gradual reduction in net rate of
replication as diffusion of the resource increasingly limits colony
growth with time.  With no additional fitting, the model also
correctly {\em predicted} the non-monotonic, upside-down U shaped
dependence of the population size on the inoculating bacterial
density. Moreover, all of the best-fit parameters inferred from the
data agreed with prior estimates in the literature, where these are
available (see Table \ref{fittedpars} and references therein),
indicating high-quality fits without overfitting.

Our study has revealed and/or confirmed several intriguing
observations about bacteria growing in colonies. First, the growth in
colonies yielded substantially greater viable cell densities than
obtained in liquid culture with the same concentrations of limiting
carbon source. We propose that this is a direct consequence of the
diffusion-limited growth, which happens at a slower division rate. In
turn, slow division is correlated with smaller size of bacterial cells
\cite{cellsize}, resulting in more bacteria for the same nutrient
amount. This slowing down is very important phenotypically ---
according to our model, over $90\%$ of all bacteria in the colony are
formed at such decreased growth rate, and the yield $a_{\rm c}$ is an
average over yields at different stages of the slowing.  We will
explore the relation between the division rate and the cell size
experimentally in a future publication. A natural extension of our
model would come from measuring the dependence of the yield on the
growth rate and then verifying if both the liquid and the colony
growth can be described by the same dependence $a=a(g(\rho))$.

Our second intriguing observation, which is supported by two
independent sets of measurements, is that the packing density inside
colonies is very low, $ \mu\sim 0.03$ CFU/$\mu$m$^3$, so that the vast
majority of a volume of a colony is not occupied by viable cells. It
is possible that the colonies are, indeed, largely void of viable
cells, with extracellular fluids and matrix fibers filling in the
gaps. Another possibility is that cells deep inside the colony are
dead or dormant due to the absence of nutrients, or due to other
effects, such as mechanical stresses, so that the viable cells that we
measure are a minority of all the bacterial cells that existed. Our
experiments show no evidence for such deviations from the minimal
growth model, but it is clear that additional studies, including
direct imaging of the colony structure, must be done in the future.

One interpretation of the close fit between the predictions of this
minimal model and the results of our soft-agar experiments is that
heterogeneities beyond nutrient access contribute little to the growth
dynamics of bacteria in colonies. It remains to be tested how general
this result is. Is the {\em E.~coli} in glucose-limited minimal medium
used in this experiment exceptional?  Will the results hold for other
bacterial species and for complex media, like broth? We propose that
the minimal model developed here be used as a baseline to address such
question of generality with other bacteria and media. Models are most
useful when they do not fit data and thus point to other factors
contributing to the studied dynamics.  For growth of bacteria in
colonies, such factors can be mechanical or other stresses, cell-cell
interactions, and others. From an evolutionary perspective
particularly intriguing in this regard would be studies of growth of
bacteria in colonies initiated with multiple cells of different
genotypes (or even species), where deviations from the model could
signal such important phenomena as clonal competition or cooperation
within a clone.

\section*{Materials and Methods}

\subsection*{Bacteria}

We used {\em Escherichia coli B, ara rpsL T6 r-m-}, originally
obtained from Seymour Lederberg \cite{Levin1972}. We employ one of the
evolved strains, REL1976, generously provided by Richard Lenski
\cite{lenski1991long}.

\subsection*{Media, culture and sampling procedures}
Overnight cultures were grown in Lysogeny Broth (LB), Becton Dickinson
(Franklin Lakes, NJ, USA), diluted in $0.85\%$ saline and introduced
into in liquid or into $0.35\%$ Bacto agar with Davis minimal
salts\cite{carlton1981gene} supplemented with $0.2 {\rm mg}/{\rm ml}$
glucose as the sole and limiting carbon source.  The liquid cultures
were maintained with 10 ml of the medium in 50 ml flasks.  For the 3-d
colony experiments, $3 {\rm ml}$ of bacteria suspended in soft agar
were put into the wells of 6-well Costar Macrotiter plates, set in a
tray with distilled water to reduce the rate of evaporation.

\paragraph*{Sampling}
The viable cell densities in both the liquid and soft agar cultures
were estimated by serial dilution in $0.85\%$ saline and plating on LB
agar.  The bacteria from the soft agar in each well were taken up with
Samco Scientific long transfer pipettes, VWR International,
transferred to glass tubes with 10 ml of $0.85\%$ saline.  To suspend
the bacteria in the soft agar, tubes with the 10 ml of saline and 3 ml
of soft agar were disrupted with the long transfer pipettes until the
soft agar was fully mixed with saline, and then vortexed for 30
seconds.

\subsection*{Numerical solution of the model}

The well-mixed Monod model, Eqs.~\eqref{qg2}, \eqref{qg3}, was solve
using {\tt ode15s} MatLab routine.  To solve the growth equations
Eqs.~\eqref{qg4}-\eqref{nrt} numerically, we rewrote the equations in
spherically symmetric coordinates, and then discretize the space into
concentric shells so that the partial differential equations become
sets of coupled ordinary differential equations describing dynamics
within each shell. These were then solved again using Matlab's {\tt
  ode15s}, with an additional constraint that redistributed the total
number of bacteria $N$ into a bacterial colony with the constant cell
packing density, as in Eq.~\eqref{nrt}, at every time step. That is,
each discretized shell of the space had a maximum cellular capacity
given by the packing density and the shell volume. The constraint
redistributed those cells that overflowed each shell's capacity to the
colony's edge, but we did not shrink the inner shells when the cells
in them started dying. Newly grown cells are first filled in the
colony's current edge shell. If the current edge shell is overflowed,
the extra cells are filled in the next shell, and so on.

\subsection*{Model fitting and confidence intervals estimation}
We first fitted the five parameters of the Monod model for the growth
in the liquid culture, Eqs.~\eqref{qg2}, \eqref{qg3}. For this we
defined the loss function
${\cal L}=\sum_i (N_i-N(t_i;\gm,m,K,\tl,a_{\rm l}))^2$, where $N_i$
was the population size (in CFU/ml) in the $i$'th measurement, and
$N(t_i;\gm,m,K,\tl,a_{\rm l})$ was the model prediction for the same
time and for given parameter values. Note that we did not average
measurements at the same $t$, but incorporated all individual
observations into the loss function, cf.~Fig.~\ref{80cband}. We
optimized ${\cal L}$ over the five parameters using MatLab's {\tt
  fmincon}. For $K$ and $m$, which are small and have large
uncertainties, we optimized w.~r.~t.\ their logarithms, thus enforcing
their positivity (the other parameters were sufficiently constrained
by data away from zero even without this reparameterization).  The
optimization was performed with ten different random initial
conditions for the parameters, and the best values from among all the
runs were chosen, resulting in the best-fit parameters
$\bar{g}_{\rm max},\bar{m},\bar{K},\bar{\tau}_{\rm l},\bar{a}_{\rm
  l}$, which we report in Tbl.~\ref{fittedpars}.

To estimate the confidence intervals for these inferences, we
bootstrapped the data 1000 times \cite{bootstrap}. When re-sampling
with replacements for bootstrapping, we resampled separately from the
exponential growth region ($t\le22$ hrs) and the saturated region
($t>22$ hrs), so that the number of data points in each of the regions
was fixed in all resampled datasets. We refitted the five growth
parameters for each of the resampled data sets. The middle 80\% of the
best-fit parameter realizations are reported in Tbl.~\ref{fittedpars}
as confidence intervals, and the covariances among the bootstrapped
best-fit values are reported in Tbl.~\ref{covariance}. Since the
sensitivities to the parameters vary widely, and ${\cal L}$ near its
minimum is badly approximated by a quadratic form, we additionally
report confidence intervals directly on the model predictions, rather
than just the parameters. For this, for each of the 1000 resampled
datasets, we calculated the population growth with the best-fit
parameters, and the middle 80\% of these growth curves are shown as the
colored band in Fig.~\ref{80cband} (top).

For fitting the 3-d growth model, Eqs.~\eqref{qg4}-\eqref{nrt}, we
write the loss function
${\cal L}=\sum_i (N_i-N(t_i;\bar{g}_{\rm
  max},\bar{m},\bar{K},\bar{\tau}_{\rm l},a_{\rm c},D,\mu))^2$.
This is minimized as above over $a_{\rm c},D,\mu$, with the first four
parameters inherited from the optimizations for liquid data. Results
of the optimization are shown in Fig.~\ref{80cband} (bottom). To
establish confidence intervals, we bootstrap the entire analysis
pipeline 30 times (the number is limited since parameter optimizations
for PDEs describing the nutrient dynamics are computationally costly),
resampling both the liquid and the 3-d colony data. While resampling
the colony data, we keep the number of data points in each of the
three regions constant (exponential, $t<24$, diffusion-limited,
$24 \le t<48$, and saturated, $t\geq 48$).  Confidence intervals on
parameters and model predictions in Fig.~\ref{80cband} (bottom) and
Fig.~\ref{multiseed} are then done as explained above.  We use the
same bootstrapped data sets to estimate the covariances and
correlations of the parameters (Tbl.~\ref{covariance}). These are
evaluated as empirical covariances and correlation coefficients of the
best-fit values for the bootstrapped data sets.

\section*{Supporting Information}


\begin{table}[!ht]
\begin{adjustwidth}{-2.25in}{0in} 
\centering
\caption{
{\bf Covariances and correlations of the fitted
    parameters.}}
\begin{tabular}{cccccccc }
$g_{\rm max}$, ${\rm hr}^{-1}$
&$K$, ${\rm \mu g}/{\rm L}$
&$a_{\rm l}$, $10^{6}$ CFU/$\mu$g\ glucose
&${\rm lag}$, ${\rm hr}$
&$m$, ${\rm hr^{-1}}$ 
&$a_{\rm c}$, ${\rm hr}$
&$D$, $10^{6}\mu {\rm m}^{2}/{\rm hr}$
&$\mu$, ${\rm CFUs}/\mu{\rm m}^3$\\
\hline
0.029  & 0.19 & 0.0018 & 0.0041 & 0.0047 & 6.0$\times 10^4$ & 0.018 & 5.1$\times 10^4$ \\
 {\em 0.99} & 3.4 & 0.022 & 0.047 & 0.23 & 0.11 & -0.018 & 0.0056 \\
 {\em 0.022} & {\em 0.015} & 0.0049 & -0.0021 & 0.061 & 0.0023 & -0.0012 & 1.0$\times 10^4$ \\
 {\em 0.26} & {\em 0.19} & {\em 0.046} & 0.041 & -0.018 & 0.0035 & -0.021 & -2.8$\times 10^4$ \\
 {\em 0.12} & {\em 0.082} & {\em 0.81} & {\em 0.50} & 1.8 & 0.070 & -0.089 & 9.1$\times 10^4$ \\
 {\em 0.021} & {\em 0.014} & {\em 0.23} & {\em 0.14} & {\em 0.28} & 0.018 & -0.012 & 4.5$\times 10^4$ \\
 {\em 0.25} & {\em 0.26} & {\em -0.046} & {\em -0.33} & {\em 0.041} & {\em -0.27} & 0.12 & 0.0033 \\
 {\em 0.19} & {\em 0.18} & {\em 0.11} & {\em -0.12} & {\em 0.24} & {\em 0.27} & {\em 0.77} & 1.5$\times 10^4$ \\
\end{tabular}

\begin{flushleft}The upper right quadrant shows in Roman font the
  covariance of the fitted parameters established by bootstrapping
  (see {\em Materials and Methods}). The diagonal are the parameter
  variances. The lower left quadrant shows the correlation coefficients
  in {\em Italic}. Units for the parameters are the same as in Table
  \ref{fittedpars}. While we report these values, we emphasize that
  these values must be interpreted with care since posterior
  distributions of the parameters are sloppy \cite{Gutenkunst:2007gl}
  and do not look like multivariate normal distributions. Instead they
  show long nonlinear ridges of parameters with nearly-equivalent
  likelihoods.
\end{flushleft}
\label{covariance}
\end{adjustwidth}
\end{table}

\section*{Acknowledgments}
This work was partially supported by the James S.~McDonnell Foundation
grant No.~220020321 (IN), by the National Science Foundation grant
No.~PoLS-1410978 (IN), and by GM098175 (BL).


\end{document}